\documentclass[aps,twocolumn,twoside,pra,showpacs,10pt]{revtex4-1}
\usepackage{mathrsfs,amsmath,amsbsy,color,graphicx,bm,amsthm,amsfonts,dsfont}

\begin{document}
\newtheorem{theorem}{Theorem}
\newtheorem{proposition}[theorem]{Proposition}
\newtheorem{corollary}[theorem]{Corollary}

\title{Strong analog classical simulation of coherent quantum dynamics}

\author{Dong-Sheng Wang}
\email{wds@phas.ubc.ca}
\affiliation{Department of Physics and Astronomy, University of British Columbia, Vancouver, Canada}

\begin{abstract}
    A strong analog classical simulation of general quantum evolution is proposed,
  which serves as a novel scheme in quantum computation and simulation.
  The scheme employs the approach of geometric quantum mechanics and quantum informational technique of quantum tomography,
  which applies broadly to cases of mixed states, nonunitary evolution, and infinite dimensional systems.
  The simulation provides an intriguing classical picture to probe quantum phenomena,
  namely, a coherent quantum dynamics can be viewed as
  a globally constrained classical Hamiltonian dynamics of a collection of coupled particles or strings.
   Efficiency analysis reveals a fundamental difference between the locality in real space and locality in Hilbert space,
  the latter enables efficient strong analog classical simulations.
  Examples are also studied to highlight the differences and gaps among various simulation methods.
\end{abstract}
\pacs{03.67.Ac, 45.20.Jj, 03.65.Yz, 03.65.Wj}
\date{\today}
\maketitle

\section{Introduction}
\label{sec:intr}

To properly understand as well as utilize quantum resources,
e.g. quantum coherence, are the main theme of modern quantum science.
Quantum coherence and its dynamics, namely, decoherence,
which essentially leads to entanglement~\cite{HHH+09},
play central roles in many studies such as
measurement and quantum-to-classical transition~\cite{Sch05},
quantum computation~\cite{NC00},
quantum resource theory~\cite{BCP14},
as well as strongly correlated many-body systems.
Besides, a more primary quest is to understand and seek the mechanism, or origin,
of quantum coherence itself.
Efforts have been made in the very early days to understand Schr\"{o}dinger equation
that describes the quantum behavior of particles in space using classical pictures~\cite{Bohm52,Fey48,Nel66,Omn92},
and many approaches prove to be significant for various applications,
such as quantum trajectories and hydrodynamics~\cite{Cha10},
and the efforts continue till nowadays.

The emerging field of quantum computation~\cite{NC00},
especially quantum simulation and classical simulation of quantum processes~\cite{BN09,BMK10}
provide new perspective to understand quantum coherence
and various quantum properties.
In this work, we raise the question whether it is possible to \emph{simulate}
(or simply put, reproduce) general quantum evolution
by classical mechanics in the spirit of quantum simulation,
and our result shows that the answer is yes.
However, it may seem impossible at first look since quantum mechanics (QM)
is well known as a generalization of classical mechanics,
as well as probability theory,
the latter two can be reached from QM through the mechanism of decoherence~\cite{Sch05}.
However, our simulation reveals that a quantum dynamics can be viewed as
a set of coupled classical dynamics with global constraints.

 Furthermore, we study the problem of whether and when such classical simulation
can be efficient, i.e., the cost of simulation scales polynomially with the size of the quantum target,
quantified by a proper measure.
Our result shows that quantum dynamics and classical (also statistical) dynamics can be described
in a unified way in terms of Hamiltonian dynamics and symplectic geometry,
however, a general quantum dynamics cannot be efficiently simulated classically.
Despite this, there are cases of practical interest that are classically tractable,
such as linear optics~\cite{AC99,KMN+07} and (discrete-time coined) quantum walk~\cite{SE12} studied in this work.
In general, classifications of quantum processes regarding classical simulation efficiencies
would be of broad implications, e.g., for complexity studies.

Our simulation of quantum dynamics is classical, analog, as well as strong, explained as follows.
In quantum simulation literature, there are many notions of simulation~\cite{BN09,BMK10,Nest11,Wan15},
and a simulation can be conveniently classified by a set of independent binary features, notably here,
digital/analog, classical/quantum, and weak/strong~\footnote{At least eight types of simulation exist,
while further classifications are surely possible in various contexts.}.
In general, digital/analog refers to whether a simulation is performed on a universal computer
or some dedicated-purpose devices specified by analog parameters
that can be mapped to those of the simulated objects~\cite{BN09},
classical/quantum refers to whether the simulator is classical or quantum,
and weak/strong, based on the weak/strong operator topology~\cite{Wan15},
refers to what the simulation is about, namely,
whether it simulates partly or completely the properties of the simulated objects.
Different simulations apply naturally in different contexts.
Classical digital simulation, e.g., to compute measurement results
on a quantum process~\footnote{Note there are many kinds of classical simulations of
quantum processes, such as a ``weak'' simulation~\cite{Nest11}, which is a sampling simulation
that samples from a probability distribution instead of computing it.},
refers to numerical simulation on classical computers (computational physics),
while quantum digital and analog simulations aim to find quantum speedup
and learn complicated quantum systems~\cite{BN09}.
Furthermore, strong simulation,
which requires the approximation of an object itself, hence a ``white box,''
is natural for quantum simulation since
quantum simulators can produce the desired quantum process itself,
such as local Hamiltonian-evolution~\cite{Llo96} and quantum channel~\cite{WBOS13,WS15} simulations.
On the contrary, weak simulation is common in classical simulation since
it only aims to yield partial information of an object,
e.g., the action of an operator on a state,
without the requirement to simulate the process itself or the process as a white box.

In the landscape of quantum simulation,
a strong analog classical (SAC) simulation,
which is suitable for probing the quantum-classical distinction,
is largely unexplored and overlooked.
Our work presents a SAC simulation of quantum evolution,
which is, on the one hand, novel in the field of quantum computation and simulation,
and, on the other hand, serves as an approach for the understanding of quantum dynamics in terms of classical pictures.
In details, the SAC simulation problem and scheme is:
given a quantum process to be simulated
(may include state preparation, evolution, final state verification, and measurement),
named as \emph{simulatee},
a procedure that employs the methods of geometric quantum mechanics and quantum tomography
designs a simulator,
which is a classical Hamiltonian system that reproduces the simulatee.
Both the simulation quality (accuracy) and cost can be precisely assessed.
Also there exist extensions and variations of the main simulation scheme.

The theory of geometric quantum mechanics (GQM)~\cite{Kibb79,Hes85,CMP90,CCM07} is employed
to construct the SAC simulation scheme,
which provides a unique viewpoint to reveal the quantum-classical distinction and connection,
e.g., in the study of geometric phase~\cite{Ber85,WLN05}.
In this work we find that efficient SAC simulations can be ensured by a locality in Hilbert space
(see the study in section~\ref{sec:exam}),
which is the notion of locality employed in the definition of quantum Turing machine~\cite{Deu85},
instead of the locality in the so-called real space,
e.g., the locality in local Hamiltonian many-body systems.
By comparison, in the framework of matrix product states~\cite{Vid03,PVWC07}
ground state properties of local Hamiltonian can be efficiently simulated on classical computers,
and the simulation cost scales with the amount of (bipartite) entanglement (or the bond dimension).
Another widely explored algebraic and computational approach is the stabilizer formalism,
and Wigner function negativity also contextuality are identified as the quantum notions that
determine the classical simulation efficiency~\cite{HWV+14,DAB+15},
which employs a weak (or sampling) simulation scheme, not the same as our SAC simulation.
Also interference, long been known as a quantum-classical distinction,
was recently identified as a resource for quantum speedup and a classical sampling simulation scheme was employed~\cite{Sta14}.
Generally speaking, the understanding of quantum-classical distinction largely depends on the simulation methods involved,
and the SAC simulation is more restrictive, hence more accurate
for the description of quantum dynamics in terms of classical pictures.
Our study also shows that the linear optics simulation of quantum computation~\cite{AC99,KMN+07}
(without nonlinear effects) can be viewed as an example of SAC simulation,
hence our work can provide an angle to reveal the connections and differences of various simulations.

In the following, section~\ref{sec:sim} develops the strong analog classical simulation framework
based on geometric quantum mechanics and quantum tomography.
Afterwards, the simulation efficiency issue is considered in section~\ref{sec:exam},
where we also analyze several practical examples and differences from other types of simulations.
In section~\ref{sec:simgen} various extensions
of the main simulation scheme are developed, including the cases for
nonunitary evolution and infinite-dimensional systems.
We conclude in section~\ref{sec:conc} with a brief summary and discussion.


\section{Simulation framework}
\label{sec:sim}

We start from finite-dimensional unitary evolution of pure states.
The generalizations to mixed states, nonunitary evolution
as well as infinite-dimensional cases are discussed later.
Quantum states live in projective Hilbert space $\mathcal{P}\mathcal{H}$
since they are normalized vectors with any global phase physically trivial.
Distance between any two states $|\psi\rangle$ and $|\phi\rangle$
is based on the overlap function $\langle \phi|\psi\rangle$.
Geometric quantum mechanics (GQM)~\cite{Kibb79,Hes85,CMP90,CCM07} shows that
the space $\mathcal{P}\mathcal{H}$ is a K\"{a}hler manifold,
which is specified by a symmetric Riemannian form, $\text{Re}(\langle \phi|\psi\rangle)$,
and a skew-symmetric symplectic form, $\text{Im}(\langle \phi|\psi\rangle)$.
The symplectic form implies that a Hamiltonian dynamics exists and
the space $\mathcal{P}\mathcal{H}$ can be viewed as a phase space.
For an orthonormal basis $\{|i\rangle\}$, a state $|\psi\rangle\in \mathcal{P}\mathcal{H}$ is mapped to a set of coefficients
$\psi_i:=\langle i|\psi\rangle$.
Name the real part $q_i:= \text{Re} (\psi_i)$ as ``position''
and imaginary part $p_i:= \text{Im} (\psi_i)$ as ``momentum''~\footnote{
Note one can also use $\psi_i$ and $\pi_i:=i \psi_i^*$ instead.
Position can be denoted by either $q$ or $x$.}
such that the normalization condition becomes
\begin{equation}\label{eq:norm}
  \langle \psi|\psi\rangle=\sum_i |\psi_i|^2=-i \sum_i \pi_i \psi_i =\sum_i (p_i^2+q_i^2) =1.
\end{equation}
As a result, the unitary evolution driven by a Hamiltonian $\hat{H}$ of pure state
\begin{equation}\label{eq:Sch}
  i |\dot{\psi}\rangle=\hat{H} |\psi\rangle
\end{equation}
can be equivalently written as Hamilton's equations
\begin{equation}\label{eq:HD}
  \frac{\partial H }{\partial q_i } =-\dot{p}_i, \;
  \frac{\partial H }{\partial p_i } = \dot{q}_i, \; \forall i,
\end{equation}
with classical Hamiltonian (or energy) $H=\langle \psi|\hat{H}|\psi\rangle$~\footnote{The classical
Hamiltonian is quadratic of the dynamical variables,
there is no higher-order terms,
which may appear for nonlinear modifications.}.
Note the above equations are equivalent to
$\frac{\partial H }{\partial \psi_i } =-\dot{\pi}_i$,
$\frac{\partial H }{\partial \pi_i } = \dot{\psi}_i, \; \forall i.$
Also the Hamilton's equations hold for time-dependent Hamiltonian $\hat{H}(t)$.
This shows that the unitary dynamics of a quantum state $|\psi\rangle$ is equivalent to
the Hamiltonian dynamics of a set of $d:=\text{dim} \mathcal{P}\mathcal{H}$ coupled ``particles'' $(q_i,p_i)$ in phase space
with the normalization condition~(\ref{eq:norm}).

The GQM above builds a close connection between QM and classical mechanics in phase space,
which provides a hidden classical picture of quantum dynamics
in terms of constrained Hamiltonian dynamics of coupled classical particles.
However, there also exist many other bases hence
other collection of hidden particles dynamics,
which are equivalent to each other via unitary basis transformations.
This is due to the extra Riemannian form for QM,
which is absent for classical case,
and related to the non-commutativity (or complementarity) of quantum operators.
This also implies that a quantum dynamics may arise from a set of Hamiltonian dynamics
such that the Riemannian form is respected.
In order to construct a SAC simulation,
the central problem is how many sets of Hamiltonian dynamics are inevitable.
It could be infinite, which turns out not to be necessary
due to the geometry of the set of quantum states.
From the information theoretic viewpoint, especially quantum tomography~\cite{NC00},
which is to reconstruct a quantum state or operation using finite number of operations,
a finite number of bases is sufficient (as long as it is complete).
The co-existence hence co-simulation of Hamiltonian dynamics in different bases
is a manifestation and also requirement of complementarity of general quantum operators.


Quantum tomography (QT), including quantum state tomography (QST) and quantum process tomography (QPT)~\cite{NC00},
is a computation process that takes an unknown quantum object (state or process) as input
and outputs its classical mathematical description, denoted by $[\cdot]$.
For instance, QST is a map
\begin{equation}\label{}
\mathcal{QST}: \mathcal{P}\mathcal{H}^{\otimes n}\rightarrow \mathbb{R}^m : |\psi\rangle^{\otimes n}\mapsto [\psi],
\end{equation}
for $n$ samples of the unknown state $|\psi\rangle$ and $[\psi]\in \mathbb{R}^m$ the description of it,
with some integers $n,m\in \mathbb{Z}^+$.
QPT is a generalization of QST that it requires to inject a complete set of input states $\{|\phi_i\rangle\}$ into
the unknown process.
The quantum-to-classical part in QT is performed by measurement,
represented by POVM $\mathcal{M}:=\{M_i\}$ such that $\sum_i M_i=\mathds{1}$
and each $M_i$ corresponds to an operation
\begin{equation}\label{}
\mathcal{M}_i:\rho\mapsto \text{tr}(\rho E_i) \geq 0.
\end{equation}
For QST of a qudit, an informationally complete measurement can be realized by
a projective measurement $\mathcal{P}$ along an operator basis $\{|\psi_i\rangle\langle\psi_i|\}$, with cardinality up to $d^2$.
Another way is to measure a complete set of operators
such that their eigenvectors contain an operator basis.
Further, QPT of a general qudit process requires in general $d^4$ measurement operations
since a QST is needed for each of the $d^2$ inputs $\{|\phi_i\rangle\}$.

Now, to build the SAC simulation, the simulator has to pass a verification test
in the spirit of quantum prover interactive proof system~\cite{ABE08}
such that a verifier, for whom the simulator is a black box,
cannot tell the simulator from the simulatee.
Here we employ QT as the verification scheme of our simulator,
that given an input state
the simulator is able to yield the correct output state.
Before we can build the simulator, we first need a SAC simulation of QT,
described as follows.

\begin{figure}
  \centering
  \includegraphics[width=.45\textwidth]{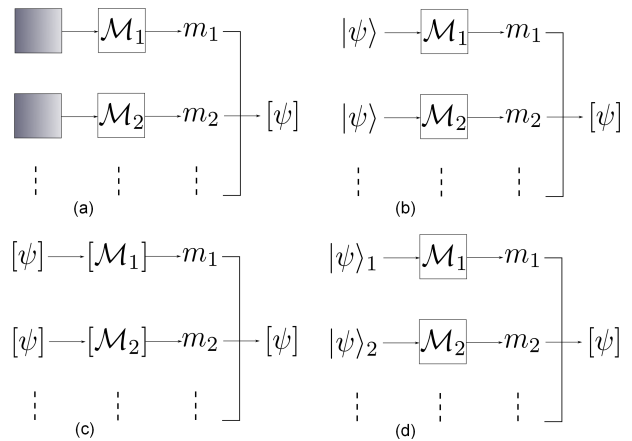}\\
  \caption{Schematic illustrations for the simulations of QST.
  (a) QST of an unknown state (the black boxes) by measurements $\mathcal{M}_i$
  with results $m_i$ that are processed into a classical computer to yield the information $[\psi]$ of the unknown state.
  (b) Quantum simulation of QST by starting from samples of the state $|\psi\rangle$.
  (c) Classical digital simulation of QST with many copies of $[\psi]$ and the information for each measurement $[\mathcal{M}_i]$.
  (d) SAC simulation of QST with fixed-basis states $|\psi\rangle_i$
  and classical measurements $\mathcal{M}_i$ of position and momentum.
  }\label{fig:SimQST}
\end{figure}

The classification of simulation also applies to the simulation of QT,
see Fig.~\ref{fig:SimQST} for the case of QST.
Note that QT, Fig.~\ref{fig:SimQST}(a), is to reveal an unknown object
while a simulation of QT requires that the object is known.
First, a quantum simulation of QST is rather straightforward
by simply performing the measurements on physical states, see Fig.~\ref{fig:SimQST}(b).
Also a classical (digital) simulation (Fig.~\ref{fig:SimQST}(c)) is
to evaluate measurement results given the information of the input quantum state,
that is, it realizes $[\psi]^{n}\mapsto [\psi] $,
which trivially simulates the behavior of QST without computing unknown quantities.
However, SAC simulation (Fig.~\ref{fig:SimQST}(d)) requires the input can indeed represent the physical state.
This is done by using a set of \emph{fixed-basis} classical states in different bases $\{\mathcal{B}_\alpha\}$;
that is, for each basis $\mathcal{B}_\alpha$,
the initial input state can be represented by a classical system (see Eq.~(\ref{eq:HD})),
and each measurement in the same basis is simulated by a classical measurement on each classical system,
which is simply to measure the position $q$ and momentum $p$ of each hidden particle~\footnote{
Note it is reasonable to require that the classical measurement result needs to be directly extractable from the classical system,
i.e., no other non-trivial computation is allowed during the measurement.}.

A precise scheme is as follows.
We use the method of measuring a complete set of operators to perform QST.
A convenient choice is a traceless orthonormal operator basis $\{\sigma_i\}$
such that $\sigma_0=\mathds{1}$, $\text{tr}\sigma_i=0$,
$\text{tr}(\sigma_i^\dagger \sigma_j)=d \delta_{ij}$.
For any state
\begin{equation}\label{eq:Bloch}
\rho =\left(  \mathds{1} + \vec{n}\cdot \vec{\sigma} \right)/d
\end{equation}
with $\vec{\sigma}:=(\sigma_i)$ and Bloch vector $\vec{n}:=(n_i)$ for $n_i=\text{tr}(\sigma_i^\dagger\rho)$
and $|\vec{n}|\leq \sqrt{d-1}$,
the measurement of $\{\sigma_i\}$ generates $\vec{n}$ that determines $\rho$.
A well-known construction is the Heisenberg-Weyl (HW) basis $\{M_{jk}\}$~\cite{BMM+12,Kim03,SZL04}
\begin{equation}\label{eq:hw}
X_j=\sum_{i=0}^{d-1}|i\rangle\langle i+j|, \quad Z_k=\sum_{l=0}^{d-1}\omega^{lk}|l\rangle\langle l| \quad (\mathrm{mod}\; d),
\end{equation}
for $M_{jk}=X_jZ_k$, and $\omega=e^{i2\pi /d}$, $j, k\in\{0,\dots, d-1\}$.
The eigenvectors of each operator $M_{jk}$ provide a complete basis,
denoted by $\mathcal{B}_{jk}:=\{|b_i\rangle_{jk},i\in \mathbb{Z}_d\}$.
A nice property is that for each fixed $j$ the set of operators $M_{jk}$ have similar eigenvectors.
Given a state $|\psi\rangle$,
QST can be done by projective measurements along each basis $\mathcal{B}_{jk}$ on different samples of the given state.
The SAC simulation of QST is to prepare up to $d^2$ classical systems,
each in a basis $\mathcal{B}_{jk}$, the corresponding state is denoted by $|\psi\rangle_{jk}$,
and then each fixed-basis state $|\psi\rangle_{jk}$ is classically measured in the basis $\mathcal{B}_{jk}$,
respectively.
The measurement results can then be programmed by a classical computer to yield $[\psi]$,
same with QST.
To simulate QPT, each input state $|\phi\rangle$ is first substituted by
the set of fixed-basis state $\{|\phi\rangle_{jk}\}$,
and the SAC simulation of QST can be done on each input.
For $d^2$ input, one has to perform $d^4$ runs of the SAC simulation.

\begin{figure}
  \centering
  \includegraphics[width=.45\textwidth]{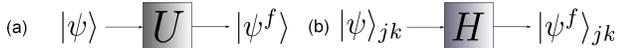}\\
  \caption{SAC simulation of a unitary evolution $U=e^{-it\hat{H}}$.
  (a) For an arbitrary initial state $|\psi\rangle$ the final state is $|\psi^f\rangle=U|\psi\rangle$.
  (b) A single run of the SAC simulator is to input a fixed-basis state $|\psi\rangle_{jk}$,
  represented by the position and momentum $(q_i,p_i)$ of $d$ hidden particles,
  and a constrained Hamiltonian dynamics with $H=\langle \psi| \hat{H}|\psi\rangle$ generates final values of $(q_i,p_i)$,
  which represents $|\psi^f\rangle_{jk}$.
  When QT verification is performed, many runs of the simulator as well as simulatee are required.
   }\label{fig:SimSAC}
\end{figure}

Now we can build the SAC simulator of a general quantum evolution based on the simulation of QT.
Given a simulatee,
the algorithm to construct the simulator is specified by:
the input is a unitary process $U$ specified by a Hamiltonian $\hat{H}$ and time $t$,
and a verification scheme specified by QT
(e.g., a complete set $\{|\phi_i\rangle\}$ for QPT and the set of bases $\{\mathcal{B}_{jk}\}$ for QST),
and the output is the simulator.
For the simulator the input states of the classical hidden particles can be obtained from QT,
the set of Hamiltonian dynamics can be obtained from QT and $U$.
To simulate an evolution $|\psi\rangle\mapsto U|\psi\rangle$,
the simulator runs the set of complementary classical hidden systems
under Hamiltonian dynamics with the normalization constraint,
and yields the final state.
If QT is employed to verify the simulator,
the simulation scheme of QT can ensure that our simulator passes the QT verification
and serves a valid simulator.
A schematic diagram is shown in Fig.~\ref{fig:SimSAC}.

 Finally, the simulation accuracy can also be quantified.
The SAC simulator can be implemented by classical point particles moving in real space,
and even also quantum particles (see subsection~\ref{subsec:loqc}).
In practice there could be many sources of simulation errors depending on the
physical implementations.
This leads to the notion of approximate simulation with an accuracy.
In this context, the distinction between weak and strong simulations becomes apparent.
Given an object $O$ and a simulation accuracy bound $\epsilon$,
strong simulation can be defined as
\begin{equation}\label{}
\sup_i \| O(i)-\tilde{O}(i) \|\leq \epsilon,
\end{equation}
while for weak simulation, the simulatee is a property of $O$, denoted as $f_O$,
and the accuracy condition is
\begin{equation}\label{}
\sup_i \| f_O(i)-f_{\tilde{O}}(i) \|\leq \epsilon,
\end{equation}
for $\|\cdot\|$ denoting a certain distance measure~\cite{Wan15}.
Also note that the supreme over all input ($i$) may not be necessary for some simulation tasks.
Now, for the SAC simulation of a unitary evolution
the simulation error can be quantified by the distance on the final state.
A good simulator should yield approximate final state given arbitrary input state.
This is the operator norm distance
\begin{equation}\label{}
  \|U-\tilde{U}\|:=\sup_{|\psi\rangle}\|(U-\tilde{U})|\psi\rangle\|
\end{equation}
for $U$ as the quantum simulatee and $\tilde{U}$ as the SAC simulator,
and $|\psi\rangle$ as input state.
The simulator may have a different input state from the simulatee
due to inaccurate initialization,
which yet does not destroy the simulation.
If the distance on the initial state is upper bounded by $\epsilon_0$,
and the distance on the evolution is upper bounded by $\epsilon$,
then the total distance is upper bounded by $\epsilon+\epsilon_0$,
following simply from the property of operator norm.

\section{Simulation efficiency and examples}
\label{sec:exam}

After establishing the primary framework of SAC simulation,
in this section we further study another important issue: the simulation efficiency,
and highlight the differences from other simulation methods by several practical examples.
We find that, different from other simulations,
the SAC simulation efficiency depends on the notions of locality:
whether it is the locality in real space (coordinate space), which is a classical notion,
or it is the locality in Hilbert space, which is a genuine quantum notion.
Here \emph{locality in Hilbert space} means that,
given an order (e.g., by eigenvalues) on a basis $\{|i\rangle\}$ of $\mathcal{H}$,
and starting with $|i\rangle$,
the dynamics acting on it only lead to a finite number of states close to it.
This notion has been implicitly used in the model of quantum Turing machine~\cite{Deu85}.
We term this type of locality as \emph{Hilbert locality} for simplicity,
different from the locality in the model of local Hamiltonian,
which is still classical.
Please note that this is also not the same as the notion of nonlocality defined by Bell's inequality.

Before we proceed, let us first extend our simulation method based on Hamiltonian dynamics to discrete-time case.
In phase space, the dynamics specified by Eq.~(\ref{eq:HD}) is symplectic,
which preserves the area defined in phase space~\cite{KN91}.
The Jacobian $J$ defined by
\begin{equation}\label{eq:Jacobian}
  J_{ij}=\frac{\partial y_i}{\partial \tilde{y}_i}
\end{equation}
for $y_i$ denoting each input variables $p_1,\dots,p_n$, $q_1,\dots,q_n$,
and $\tilde{y}_i$ denoting each output variables
$\tilde{p}_1,\dots,\tilde{p}_n$, $\tilde{q}_1,\dots,\tilde{q}_n$
preserves the symplectic form
\begin{equation}\label{}
  \Delta=\begin{pmatrix}
    0 & -\mathds{1} \\ \mathds{1} & 0
  \end{pmatrix}
\end{equation}
such that $J\Delta J^t= \Delta$.
In order to consider the simulation of circuits and algorithms in quantum computation,
which often involve discrete-time execution of a sequence of gates,
we also allow symplectic matrix $S\in Sp(2n,\mathbb{R})$ in our SAC simulation.
As both continuous-time and discrete-time evolution are common in reality,
we allow both continuous-time and discrete-time types SAC simulation,
similar with the case of quantum simulation~\cite{Llo96,SE12,WBOS13,WS15},
e.g., there are both continuous-time and discrete-time quantum walks.

In the following we study several examples,
including linear optics, quantum walk,
local Hamiltonian evolution,
stabilizer circuits,
and matrix product states to reveal the main features of SAC simulation.
For simplicity, we ignore the verification part by tomography of the simulation
and the study of simulation accuracy,
while we focus on the efficiency of SAC simulations and differences with other
simulation methods.

\subsection{Hilbert locality: Linear optical quantum computation}
\label{subsec:loqc}

\begin{figure}[b!]
  \centering
  \includegraphics[width=.15\textwidth]{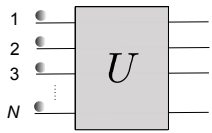}\\
  \caption{Schematic diagram for linear optical quantum computation of a unitary $U\in SU(N)$
  built by beam splitters and phase shifters~\cite{RZB+94} acting on input photons with $N$ spatial modes.}\label{fig:loqc}
\end{figure}

In this section we study a nontrivial setup that can be thought of as an efficient SAC simulator
that benefits from Hilbert locality.
There are many different approaches for using linear (and also nonlinear) quantum optics
for quantum computation, here we analyze the approach using the so-called dual-rail encoding~\cite{AC99,KMN+07}.
As illustrated in Fig.~\ref{fig:loqc}, to simulate a unitary gate $U\in SU(N)$,
$N$ paths (or called modes) are employed to represent the $N$ levels of the input and output system,
and single photon can be injected into each mode.
The unitary $U$ can be built up by $O(N^2)$ beam splitters and phase shifters~\cite{RZB+94}.
It's clear to see that if $N=2^m$ for $m$ as the number of qubits,
the simulation of an $m$-qubit unitary by such linear optics setup is not efficient.

The unitary gate $U$ is usually understood in terms of the mapping
\begin{equation}\label{}
  U: (\hat{a}^\dagger_1,\dots, \hat{a}^\dagger_N)\mapsto (\hat{b}^\dagger_1,\dots, \hat{b}^\dagger_N)
\end{equation}
for input (output) creation operator $\hat{a}^\dagger_i$ ($\hat{b}^\dagger_i$).
Note that the photon in each mode has trivial self-evolution,
and the dynamics on all the photons are driven by external optical elements.
As $\hat{a}^\dagger_i=\hat{q}_i+i \hat{p}_i$,
let $U=V+iW$, then it is equivalent to the symplectic matrix
\begin{equation}\label{}
  S_U:=\begin{pmatrix}
    V & -W \\ W & V
  \end{pmatrix}\in Sp(2N,\mathbb{R})
\end{equation}
acting on $\hat{q}_1,\dots,\hat{q}_N,\hat{p}_1,\dots,\hat{p}_N$.

This linear optics setting can perform small-scale efficient simulation of quantum circuits on qubits
and also local Hamiltonian evolution~\cite{KMN+07}.
The simulation is strong since it simulates the process itself instead of observable effects,
and is classical in the sense that the dynamics can be understood in the phase space picture.
However, it is not totally analog, instead it is digital
since it uses universal elements (beam splitter and phase shifter) to represent the simulatee.
Furthermore, the linear optics simulation is even not totally classical since the photon space is
actually second-quantized. That is, the space of the input photons is not
\begin{equation}\label{}
  \mathbb{C}^N=\mathbb{C}\oplus \mathbb{C} \oplus \cdots \oplus \mathbb{C},
\end{equation}
instead, each mode $\mathbb{C}$ is second-quantized to a Fock space $L_2(\mathbb{R})$,
which is acted upon by $\hat{q}_i$ and $\hat{p}_i$.
The unitary $U$ is equivalent to another unitary $\mathbb{W}$,
the so-called metaplectic representation~\cite{Hol11},
acting on $L_2(\mathbb{R}^N)$.
This second-quantization feature goes beyond the phase space framework in this paper
(yet, see subsection~\ref{subsec:inf} for a further analysis),
which is still a classical space,
and it benefits the linear optics simulation for other tasks, e.g., notably,
the boson sampling algorithm~\cite{AA11}.

However, we can still interpret the linear optics simulation using the dual-rail encoding
as a SAC simulation if the second-quantization feature is simply dismissed,
and the simulation of $U$ can also be understood as the simulation of a continuous-time evolution
for $U=e^{-it\hat{H}}$ with a Hamiltonian $\hat{H}$.
As a result, the linear optics setup serves as an example of SAC simulation in the generalized sense,
and the photon in each mode can be viewed as the hidden particle in our terminology.
The merit of the optics setup is that
the optical elements (beam splitters and phase shifters) are local operations in Hilbert space,
hence efficient simulation can be built up as long as the total dimension $N$ does not scale
exponentially with the problem size, e.g., the number of qubits $m$ of the simulatee.

\subsection{Hilbert locality: Discrete-time quantum walk}
\label{subsec:qwalk}

Besides linear optics, another model that employs the Hilbert locality is
the discrete-time coined quantum walk~\cite{SE12},
which has been proven to be a universal model for quantum computing.
It can be viewed as a simplified version of a quantum Turing machine~\cite{Deu85}
so that the state of the walker specifies the register tape
and the state of the coin specifies the control processor.
Here we analyze the standard setting of quantum walk from the viewpoint of SAC simulation,
we find that a SAC simulation of quantum walk is efficient.

In coined quantum walk, each step is specified by
\begin{equation}\label{eq:qwalk}
  U=S(H\otimes \mathds{1})
\end{equation}
for Hadamard $H$ acting on the coin qubit and
$S$ a uniformly-controlled shift operator
\begin{equation}\label{}
  S=P_0\otimes X^\dagger + P_1\otimes X
\end{equation}
with the coin as control and the walker as target,
and the shift operator
\begin{equation}\label{eq:shift}
X=\sum_{x=-d}^d |x\rangle\langle x+1|
\end{equation}
is a generator of the Heisenberg-Weyl group (see Eq.~(\ref{eq:hw}))
and the walker space dimension is $2d+1$.
Then the system (c$\otimes$w) starting from a product state
$  |\psi(0)\rangle=|\psi(0)\rangle_c |0\rangle_w$
evolves after $T$ (for $d\geq T$) steps and yields
$  |\psi(T)\rangle=(U)^T |\psi(0)\rangle.$
The mixing time evaluated from the final probability distribution
$\mathcal{P}(x)$ of the walker position is $O(T)$,
which shows a quadratic speedup than the classical random walk,
which is $O(T^2)$.

\begin{figure}[t!]
  \centering
  \includegraphics[width=.48\textwidth]{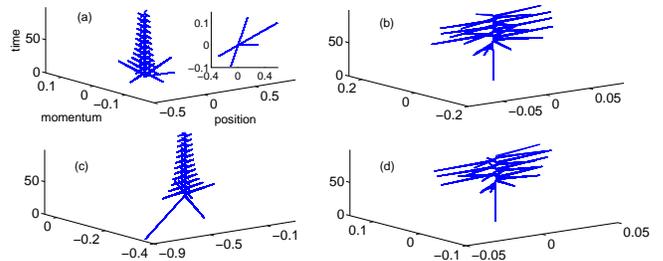}\\
  \caption{The SAC simulation of coined quantum walk for step $d=T=100$ in phase space
  for mode $|0\rangle_c|x=0\rangle_w$ (a), $|1\rangle_c|x=0\rangle_w$ (c),
  $|0\rangle_c|x=40\rangle_w$ (b), and $|1\rangle_c|x=40\rangle_w$ (d).
  The initial state is the product of a random $|\psi(0)\rangle_c$ and $|0\rangle_w$.
  All other modes also show similar trajectories.
  Insert panel in (a) is a top view showing clearly the `equilibrium point' $(0,0)$,
  which is the same for all other modes.
   }\label{fig:qwalk}
\end{figure}

For SAC simulation of this model, indeed the linear optics setup discussed above
can be employed, which has also been realized in practice.
The shift operator $X$~(\ref{eq:shift}) is two-local in Hilbert space
and the operator $S$ is also local, hence each step can be efficiently simulated,
and the total simulation cost scales linearly with $T$.
In particular, in SAC simulation the evolution of each mode $x(t)$ is primary and
can reveal more information of the algorithm
than the final probability distribution $\mathcal{P}(x)$.
In Fig.~\ref{fig:qwalk} we simulated the evolution of several modes in phase space for $d=T=100$
and a randomly chosen coin qubit state.
The phase space behavior shows that each mode (hidden particle) undergoes peculiar
non-stationary dynamics,
and the hidden particles interact with each other.
The particle at the center ($x=0$) decays in a choppy way to the `equilibrium point' $(0,0)$ in phase space,
while the particle at $x=40$ starts to oscillate at a later time.
In other words, those particles form the media for the propagation of the input wave packet,
which spreads out with a linear mixing time.
On the contrary, there is no such phase space dynamics and a set of hidden particles
for the classical random walk model,
which shows a fundamental difference between the classical and quantum cases.
However, the classical case can be achieved when there exists significant decoherence~\cite{Ken06},
and a generalized SAC simulation involving random bits can be developed (see Sec.~\ref{subsec:mix}),
which would not be analyzed explicitly here.

\subsection{Separation between SAC simulation and quantum simulation}
\label{subsec:mutip}

In this subsection we analyze the simulation scheme in the multipartite setting
when the system contains several physically local subsystems.
This serves as an example that allows efficient quantum simulation~\cite{Llo96}
yet does not allow efficient SAC simulation,
and also as an example with physical locality in the real space but not the Hilbert locality.

A quantum many-body system is usually described by a local Hamiltonian,
which is a sum of local terms, each of which only acts on a small portion of the whole system.
To be precise, an $n$-particle $k$-local Hamiltonian $\hat{H}=\sum_{\lambda=1}^{\Lambda} \hat{H}_{\lambda}$
is a sum of terms $\hat{H}_{\lambda}$ which acts on at most $k$ particles,
with $\Lambda\in O(n^k)$.
A standard quantum simulation problem is to simulate the unitary evolution
$U=e^{-it\hat{H}}$ for a given time $t$~\cite{Llo96}.
Now a SAC simulation of a local Hamiltonian evolution $U$ can be constructed as follows.
First, due to the non-commutability of the local terms,
Trotter-Suzuki formula~\cite{Tro59,Suz90} is employed to approximate $U$ by
$\tilde{U}=[U_{\chi}(\tau)]^r$, with $t=r\tau$ and $s=1/(4-4^{1/(2p-1)})$, $1<p\le \chi$,
  \begin{equation}
    U_1(\tau)=\prod_{\lambda=1}^{\Lambda} U_{\lambda}(\tau/2)\prod^1_{\lambda=\Lambda}U_{\lambda}(\tau/2),
  \end{equation}
  \begin{equation}
    U_p(\tau)=\left[U_{p-1}\left(s\tau\right)\right]^2U_{p-1}\left((1-4s)\tau\right)
			\left[U_{p-1}\left(s\tau\right)\right]^2,
  \end{equation}
and $U_{\lambda}(\tau/2)=e^{-i\hat{H}_{\lambda}\tau/2}$.
The error of this approximation is the operator-norm distance
$\|U-\tilde{U}\|\in O\left(t^{2\chi+1}/r^{2\chi}\right)$,
better than $O(t^2)$ if Trotter's formula is employed~\cite{Tro59}.
There is a tradeoff between the approximation error and the number of gates:
the error decreases as $\chi$ increases,
while there is a number of gates exponential with $\chi$ in the sequence.
As a result, in practice an optimal value of $\chi$ can be chosen for one's purpose.
Then the given local Hamiltonian evolution $U$ is formally expressed as
\begin{equation}\label{eq:useq}
U\approx \tilde{U}=U_1U_2\cdots U_m
\end{equation}
with each unitary specified by a local term
$U_i=e^{-it_i \hat{H}_\lambda}$ and a corresponding time interval $t_i$.
Hence the simulation of $U$ can be reduced to the simulation of each $U_i$,
which can be simulated by a Hamiltonian dynamics given $t_i$ and $\hat{H}_\lambda$.
Given a basis $\{|{\bf i}\rangle:=|i_1,i_2,\dots,i_n\rangle\}$ of an $n$-partite system,
if each local dimension is $d$,
there will be $d^n$ pair of variables $(q_{{\bf i}}=\text{Re}(\psi_{{\bf i}}),
p_{{\bf i}}=\text{Im}(\psi_{{\bf i}}))$
for a state $|\psi\rangle=\sum_{{\bf i}} \psi_{{\bf i}}|{\bf i}\rangle$.
Note that here the position $q_{{\bf i}}$ and momentum $p_{{\bf i}}$ do not belong to any of the local particles.
That is, $(q_{{\bf i}}, p_{{\bf i}})$ are not variables of real particles,
instead they describe hidden particles for each basis state $|{\bf i}\rangle$ of the whole Hilbert space.
Also all the hidden particles $(q_{{\bf i}}, p_{{\bf i}})$ participate in
each local Hamiltonian dynamics driven by $H_\lambda=\langle\psi| \hat{H}_\lambda|\psi\rangle$.
This means that although each term $\hat{H}_\lambda$ is local
in the sense that it acts on a limited number of locally connected particles,
the action of $\hat{H}_\lambda$ is globally on the whole Hilbert space,
hence on all hidden particles $(q_{{\bf i}}, p_{{\bf i}})$.
Finally, the SAC simulation composes a sequence of Hamiltonian dynamics corresponding to
the sequence~(\ref{eq:useq}).

We can see that although there are physically separable subsystems,
the global entanglement among them,
as well as the global action of each term $U_i=e^{-it_i \hat{H}_\lambda}$ on all hidden particles
require the system to be treated as a whole.
This contrasts with classical systems that allow classical correlations but no entanglement,
which is a quantum global constraint.
The dimension of the space of $n$ quantum particles grows exponentially with $n$ due to entanglement,
while the dimension of the phase space of $n$ classical particles only grows linearly with $n$.

\subsection{Separation between SAC simulation and classical digital simulation}
\label{subsec:mps}

\begin{figure}[t!]
  \centering
  \includegraphics[width=.35\textwidth]{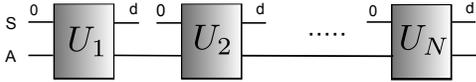}\\
  \caption{Representation of a matrix product state
  for system S that contains $N$ qudits.
  Each qudit is initialized at $|0\rangle$,
  the trace on which leads to a channel $\mathcal{E}_n$
  on the ancilla A, which is dilated to a unitary $U_n$.
  A single qudit gate affects a local $U_n$,
  while a two-qudit gate on qudits $n$ and $n+1$
  affects $U_n$ and $U_{n+1}$.
  }\label{fig:mps}
\end{figure}

Further, we analyze two examples to illustrate the differences
between SAC simulation and common simulation methods on classical (digital) computers.
We find that the previously claimed efficiencies do not hold for SAC simulation,
which manifests that our SAC simulations assess the simulation cost in a more complete way.

Consider the computation on a matrix product state (MPS)~\cite{PVWC07} that takes the form
\begin{equation}\label{eq:mps}
  |\Psi\rangle=\sum_{i_1\cdots i_N} \langle R| A(i_N)\cdots A(i_1) |L\rangle |i_1 \cdots i_N\rangle,
\end{equation}
which can be prepared by an ancilla-driven sequential quantum circuit~\cite{SSV+05},
see Fig.~\ref{fig:mps}.
Each set of local tensors $\{A(i_n)\}$ defines a channel $\mathcal{E}_n$ acting on the ancilla,
the dimension of which is called the bond dimension $\chi$,
which can vary from site to site yet usually assumed as the largest one.
It is clear to see that
the generation of an MPS on a quantum computer is efficient as long as the bond dimension
does not scale exponentially with the number of qudits in the system.
Also the (digital) simulation on classical computer is efficient since it does not contain
exponential number of parameters, instead, it is $O(\chi^2 dN)$.
However, it turns out this generation process does not permit efficient SAC simulation.
Although each gate $U_n$ is local, namely, acting on the ancilla and one qudit,
its effect is on all other qudits before it since they are entangled together,
i.e., it has nonlocal effects in the Hilbert space.

Furthermore, it has been known that~\cite{Vid03} a polynomial circuit with at most two-local gates
on an MPS can be efficiently simulated on classical computers such that
the bond dimension at each stage does not blow up.
The effects of gates can be efficiently updated on the local tensors,
and the whole circuit can be described as a sequence of MPS with small bond dimensions.
On the contrary, for similar reasons,
it is straightforward to see that
the SAC simulation of such computation cannot be efficient
since the local gates do not enjoy Hilbert locality.
Also it should be pointed out that classical simulation is usually weak simulation
since often merely some dominant observable effects (after a process) are concerned.
With the MPS form, the properties of the system can be expressed in terms of the properties of the ancilla
(e.g., the channels $\{\mathcal{E}_n\}$).
But strong simulation of a process is not designed for particular observable effects~\cite{Wan15},
which is also a feature of SAC simulation
and serves as a difference from classical digital simulation.

Besides the MPS formalism for efficient description of states with local structures,
another widely studied formalism is the stabilizer states.
In quantum computing the Gottesman-Knill theorem~\cite{NC00} shows that stabilizer circuits
can be efficiently simulated on classical computers.
The reason is that in a stabilizer circuit the state at each stage can be specified by its
set of stabilizers, which is an efficient description,
and then a quantum computational outcome can be simulated by analyzing the stabilizers.
In our notation, the description (or representation) $[\psi]$ of a stabilizer state $|\psi\rangle$ is efficient,
while by comparison, a general state needs an exponential number of parameters for its description.
In the stabilizer formalism, even highly entangled states,
such as cluster states~\cite{BR01} can be efficiently simulated.

In SAC simulation, however, it is not only required an efficient \emph{description},
instead it requires to use classical systems to mimic or reproduce the actual quantum simulatee,
which is much more stronger than the requirement of a description.
To simulate an $n$-qubit stabilizer circuit, if there exists global entanglement on all the qubits,
the simulation cannot be efficient.
For instance, consider the generation of a linear cluster state and action of gates from the Clifford group
and Pauli observable measurements on it.
In the MPS form~(\ref{eq:mps}) a linear cluster state has bond dimension $\chi=2$
and on-site tensors $H$ and $HZ$,
while in stabilizer form its stabilizers take the simple form $ZXZ$,
hence the digital simulation of the generation process is efficient,
either due to its small bond dimension or the simple stabilizer description.
However, generically as the cluster grows the number of terms in the expansion of the state grows exponentially~\cite{BR01},
which means, for SAC simulation, exponential number of hidden particles are required.
Furthermore, for gates and measurements,
the digital simulation is efficient since it is still within the stabilizer formalism,
while SAC simulation is not efficient since Clifford gates on a local qubit do not possess Hilbert locality,
similar with the case studied in subsection~\ref{subsec:mutip}.
As a result, the two examples above demonstrate that SAC simulation
is a more restrictive method than the common simulation methods on classical computers,
and an efficiency gap between them exists.


\section{Extensions of the simulation scheme}
\label{sec:simgen}

In this section we further consider several extensions of the SAC simulation scheme,
namely, to the cases of nonunitary evolution and infinite-dimensional system.
We find that these generalizations can be properly achieved.

\subsection{Mixed state and nonunitary evolution}
\label{subsec:mix}

The generalization to mixed state is straightforward,
while different methods are available depending on various decompositions of mixed state.
In this subsection we provide two different methods.
First, if one interprets $\rho$ as a convex mixture of several pure states $\{p_i, |\psi_i\rangle\}$,
then the only extra resource is a classical random number to generate the probability distribution $\{p_i\}$,
and then an average is taken over the set of evolution.
Along with the mixture form of states,
the way to handle nonunitary evolution,
described as completely positive trace-preserving mappings~\cite{Cho75}, i.e. channels,
is to employ dilation method
to convert a given channel into a unitary evolution acting on a bigger space
formed by the system and an ancilla with initial state $|0\rangle$ such that

\begin{equation}\label{eq:channel}
  \mathcal{E}(\rho) \mapsto \text{tr}_\text{A} \left( U (\rho\otimes |0\rangle\langle 0|)U^\dagger \right),
\end{equation}

\noindent and $\text{tr}_\text{A}$ represents the trace of the ancilla,
which can be realized by a projective measurement along an ancillary basis $\{|i\rangle\}$.
A mixed system input state $\rho$ can be viewed as a mixture of several pure states $\{p_i,|\psi_i\rangle\}$,
hence the dynamics~(\ref{eq:channel}) can be simplified as a mixture of $\mathcal{E}(|\psi_i\rangle)$
with pure state input.
The trace operation can also be simulated statistically:
for each projection $|j\rangle\langle j|$, the resulting system state is $K_j|\psi_i\rangle$ (up to renormalization)
for Kraus operator $K_j:=\langle j|U|0\rangle$
with probability $q_{ij}:=\langle\psi_i|K_j^\dagger K_j|\psi_i\rangle$.
As a result, the SAC simulator can be constructed as a mixture, according to $\{p_i\}$ and $\{q_{ij}\}$,
of the simulation for each pure system state input and each projective operation on the ancilla.
The evolution $U$ can be simulated by Hamiltonian dynamics,
and the simulation accuracy is quantified by distance on channels~\cite{WS15}.
In addition, this simulation scheme also highlights the feature of quantum mechanics compared with
classical mechanics and statistical mechanics:
if a general quantum state $\rho$ is treated as a mixture of pure states $|\psi_i\rangle$
with corresponding probabilities $p_i$,
probability theory captures the part of quantum dynamics specified by $p_i$, ignoring the details of each state $|\psi_i\rangle$,
while classical mechanics captures the part of quantum dynamics of each state $|\psi_i\rangle$,
ignoring the statistics specified by $p_i$,
and quantum theory as a whole is a consistent combination of them.

Second, we propose another scheme that does not consume classical random numbers.
According to channel-state duality~\cite{Cho75},
a quantum channel $\mathcal{E}$ can be equivalently represented as a state,
the so-called Choi state
$\mathcal{C}=\mathcal{E}\otimes \mathds{1} (\eta)$,
for $\eta=|\eta\rangle\langle \eta|$ and $|\eta\rangle=\sum_i |ii\rangle$.
For example, a unitary operator $U=(u_{ij})$ can be represented by a vector $|U\rangle=\sum_{ij}u_{ij}|ij\rangle$,
which is actually the reshaping of $U$.
Also a state $\rho=(\rho_{ij})$ is represented by $|\rho\rangle=\sum_{ij}\rho_{ij}|ij\rangle$.
The inner product of any two operators $A$ and $B$ is $\text{tr}(A^\dagger B)=\langle A|B\rangle$.
Now Eq.~(\ref{eq:Bloch}) can be equivalently written as
\begin{equation}\label{eq:Bloch2}
  |\rho\rangle=(|\eta\rangle +\sum_i n_i |\sigma_i\rangle)/d
\end{equation}
for $|\eta\rangle=|\mathds{1}\rangle=|\sigma_0\rangle$,
and in general, $n_i\in \mathbb{C}$, e.g., in HW basis.
This is the expansion of $|\rho\rangle$ in the basis $\{|\sigma_i\rangle\}$.
Let $n_0=1$, $\mathscr{Q}_i:=\text{Re}(n_i)$, $\mathscr{P}_i:=\text{Im}(n_i)$,
then a state $\rho$ can be treated as a set of $d^2$ classical hidden particles
$(\mathscr{Q}_i, \mathscr{P}_i)$ as in the pure state case.
Also now the normalization is
\begin{equation}\label{eq:norm2}
\langle\rho|\rho\rangle=\text{tr}\rho^2=\frac{1}{d}(1+|\vec{n}|^2)=\frac{1}{d} \sum_i (\mathscr{Q}_i^2 + \mathscr{P}_i^2) .
\end{equation}
We can say that the mixed state case is ``second order'' and pure state case is ``first order''
due to their definitions of hidden particles and
normalization conditions (higher-order states can also be constructed in the framework of superchannel~\cite{CDP08}).
Now the unitary dynamics $i\dot{\rho}=[\hat{H},\rho]$ is mapped to
\begin{equation}\label{eq:vonN}
  i|\dot{\rho}\rangle=\mathscr{\hat{H}}|\rho\rangle
\end{equation}
for Hamiltonian $\mathscr{\hat{H}}:=\hat{H}\otimes\mathds{1}-\mathds{1}\otimes\hat{H}^*$.
The unitary evolution $U=e^{-it\hat{H}}$ is mapped to $\mathscr{U}=U\otimes U^*$.
Note the above forms reduce to the case when $\rho$ is pure:
a state $|\psi\rangle\langle\psi|$ is mapped to $|\psi\rangle|\psi^*\rangle$,
and evolution~(\ref{eq:vonN}) reduces to Eq.~(\ref{eq:Sch}).
Defining the classical Hamiltonian as $\mathscr{H}=\langle\rho|\mathscr{\hat{H}}|\rho\rangle$,
now it is direct to find that Eq.~(\ref{eq:vonN}) can be written as
\begin{equation}\label{eq:HD2}
  \frac{\partial \mathscr{H} }{\partial \mathscr{Q}_i } =-\dot{\mathscr{P}}_i, \;
  \frac{\partial \mathscr{H} }{\partial \mathscr{P}_i } = \dot{\mathscr{Q}}_i, \; \forall i,
\end{equation}
as a generalization of~(\ref{eq:HD}).
There is a crucial difference between them:
the $d^2$ hidden particles are coupled together for the mixed state case~(\ref{eq:HD2}),
and the variables $(\mathscr{Q}_i, \mathscr{P}_i)$ are ``second order,''
while only $d$ hidden particles are coupled for the pure state case~(\ref{eq:HD})
but $d$ of them are needed to form a complementary complete set, so in total $d^2$,
and the variables $(q_i, p_i)$ are ``first order.''
A SAC simulator for the mixed state case can be built straightforwardly.

Next, for nonunitary evolution we consider Lindblad equation~\cite{Lin76}
$i \dot{\rho}=\mathcal{\hat{L}}\rho$ for
$\mathcal{L}\rho=[\hat{H},\rho]+i\sum_i\gamma_i (L_i \rho L_i^\dagger-\frac{1}{2}L_i^\dagger L_i\rho-\frac{1}{2}\rho L_i^\dagger L_i)$,
which can be equivalently written as $i|\dot{\rho}\rangle=\mathscr{\hat{L}}|\rho\rangle$
for an effective Hamiltonian
\begin{equation}\label{}
  \mathscr{\hat{L}}:=\mathscr{\hat{H}}+ i\sum_i\gamma_i ( L_i\otimes  L_i^*
  - \frac{1}{2} L_i^\dagger L_i \otimes \mathds{1}- \frac{1}{2}  \mathds{1} \otimes L_i^t  L_i^* ).
\end{equation}
Its classical version can be defined as
$\mathscr{L}=\langle\rho|\hat{\mathscr{L}}|\rho\rangle$,
which still drives a Hamiltonian dynamics, similar with~(\ref{eq:HD2}),
but now the normalization~(\ref{eq:norm2}) becomes time-dependent
since the evolution changes the Bloch vector length $|\vec{n}|$.

By comparison, the first scheme, ``mixture plus dilation,'' requires
a larger space hence the trace operation and classical bits,
but the evolution can be simulated the same as pure state unitary evolution.
While the ``second order'' vector method does not require a larger space,
but the evolution becomes more complicated, e.g.,
the normalization condition can be time-dependent.
This shows a tradeoff between them.

\subsection{Infinite dimensional cases}
\label{subsec:inf}

Finally, we show that the simulation scheme can also be generalized to infinite dimensional cases.
As expected, the dynamics is not for a collection of discrete particles,
instead it is for a field.
The finite case above can be viewed as a discretization of a field.
A state $|\psi\rangle\in L_2(\mathbb{R})$ can be expanded in
the position basis $\{|x\rangle\}$ or momentum basis $\{|p\rangle\}$ for $x,p\in \mathbb{R}$
as $|\psi\rangle= \int dx \psi(x) |x\rangle$
or $|\psi\rangle= \int dp \phi(p) |p\rangle$.
The unitary evolution now becomes
\begin{equation}\label{}
  i \dot{\psi}(x)= H(x) \psi(x),
\end{equation}
for $H(x)=\langle x|\hat{H}|x\rangle$.
In the position basis,
the corresponding ``momentum'' variable of $\psi(x)$ is $\pi(x)\equiv i\psi^*(x)$,
also we can use $\text{Re}\psi(x)$ and $\text{Im} \psi(x)$ as in the discrete case.
With energy $H=\int dx \psi^*(x) H(x) \psi(x)$, the Hamilton dynamics is
  \begin{equation}\label{}
  \frac{\partial H }{\partial \psi(x) } =-\dot{\pi}(x),
  \frac{\partial H }{\partial \pi(x) } = \dot{\psi}(x).
\end{equation}
Different from a true classical field dynamics,
here both the dynamics of $\psi(x)$ and $\phi(p)$ are needed,
which necessarily leads to the uncertainty relation on position operator $\hat{x}$ and
momentum operator $\hat{p}$.
Further, the information contained by $\psi(x)$ and $\phi(p)$
is just equivalent to the well-known Wigner function $W(x,p)$.
A SAC simulator can be simply built by the Hamiltonian dynamics
for both $\psi(x)$ and $\phi(p)$,
and tomography can be performed by technique to measure Wigner function.

 This scheme can be further generalized to cases
when there are both discrete and continuous degree of freedoms.
The Hamilton's equations then describe the dynamics of several coupled fields
  \begin{equation}\label{}
  \frac{\partial H }{\partial \psi_i({\bf x}) } =-\dot{\pi}_i({\bf x}),
  \frac{\partial H }{\partial \pi_i({\bf x}) } = \dot{\psi}_i({\bf x})
\end{equation}
for energy
\begin{equation}\label{}
H=\sum_{i=1}^N\int d{\bf x} \psi^*_i({\bf x}) H_{ij}({\bf x}) \psi_j({\bf x}),
\end{equation}
and ${\bf x}=x_1\cdots x_N$,
$H_{ij}({\bf x})=\langle i|\langle {\bf x}| \hat{\mathbb{H}}|{\bf x}\rangle|j\rangle$,
and a Hamiltonian operator $\hat{\mathbb{H}}$ acting on space $L_2(\mathbb{R}^N)$.
In fact, this can be employed to simulate the evolution of photons in the linear optics setup
studied in subsection~\ref{subsec:loqc} when the second-quantization feature of photons are considered.
The metaplectic representation $\mathbb{W}$ relates to $\hat{\mathbb{H}}$ such that
\begin{equation}\label{}
  \mathbb{W}=e^{-it \hat{\mathbb{H}}}.
\end{equation}
This shows that a second-quantized system can also be described in phase space.
However, a further analysis of nonunitary dynamics, e.g., Gaussian channels,
would go beyond the current scope of this work.

\section{Conclusion}
\label{sec:conc}

 In this work two central problems have been studied:
first, whether it is possible to simulate quantum evolution by classical means
in a stronger sense, which is formalized as strong analog classical (SAC) simulation;
second, whether such simulation can be efficient for specific quantum processes.
Our study shows that indeed quantum coherent dynamics can be described in phase space,
which is a standard framework for classical mechanics,
hence the quantum-classical distinction can be revealed by SAC simulation.
More important issue is efficiency, which is a central concept in computer science rather than physics,
and we find that the locality in Hilbert space serves as a sufficient condition for
efficient SAC simulation.

We have constructed a SAC simulation scheme
mainly for continuous-time Hamiltonian evolution with verification requirement
and analysis of simulation accuracy.
The scheme is generalizable to cover cases of discrete-time evolution,
nonunitary evolution, and infinite-dimensional systems.
Our simulation shows that quantum dynamics can be treated as a set of complicated,
but geometrically concise, coupled classical dynamics;
roughly, a bunch of complementary strings (set of particles for the discrete case) driven by Hamiltonian dynamics.
Compared with other frameworks, e.g., the theory of contextuality~\cite{HWV+14,DAB+15} or entanglement~\cite{HHH+09},
our approach is dynamical rather than kinematical or algebraic.
However, our studies do not intend to make any connection with the hidden variable theories,
instead our simulation is based on geometric quantum mechanics,
which has been a novel approach for many studies including geometric phase.

The study of simulation efficiency manifests the role of locality in quantum coherent dynamics.
Two distinct notions of locality can be separated by the SAC simulation:
dynamics with locality in real space cause problems for efficient SAC simulations,
while dynamics with locality in Hilbert space permit efficient SAC simulations.
The locality in real space is more common in physics,
yet the Hilbert locality is less understood.
The examples of linear optical quantum computing and quantum walk demonstrate
the central roles of Hilbert locality in quantum computing and quantum algorithms,
which may benefit the designs of new quantum algorithms.
As well, at present it is not clear whether Hilbert locality is necessary for SAC simulation.
The relation between these two kinds of localities is also a worthy topic by itself.

We also made the efficiency separation between the SAC simulation and other simulation methods,
including quantum simulation and classical digital simulation.
These comparison shows that the SAC simulation accounts for simulation costs more completely,
hence revealing the quantum-classical distinction more faithfully.
Efficiency of classical simulations depends on the simulatee and the involved simulation methods.
In the spirit of resource theory,
our study highlights the role of quantum entanglement,
especially global entanglement,
due to which a multipartite quantum system is preferred to be treated as a whole
instead of separable parts.

As well, our results also trigger the following question:
can the \emph{simulator} be treated as a \emph{model} of the true physical reality?
This should be examined with care since, e.g.,
the concept of spin does not really mean a particle is spinning.
However, this quests further investigation to understand what quantum coherence actually is.
In addition, it would also be interesting and important to explore other related problems,
such as other verification methods besides tomography.

\begin{acknowledgments}
The author would like to thank R. Cockett, P. H{\o}yer, and B. Sanders for the discussions
on the notion of simulation at an early stage of this work, and R. Raussendorf for comments.
Funding support from NSERC of Canada and
a research fellowship at Department of Physics and Astronomy, University of British Columbia
are acknowledged.
\end{acknowledgments}

\bibliography{oss}

\end{document}